\documentclass[twocolumn,showpacs,amsmath,amssymb,aps,prc]{revtex4}

\usepackage{graphicx}
\usepackage{dcolumn}
\usepackage{bm}

\def\be{\begin{equation}}
\def\ee{\end{equation}}
\def\bq{\begin{eqnarray}}
\def\eq{\end{eqnarray}}

\begin{document}
\title{Chemical and mechanical instability in warm and dense nuclear matter}
\author{A. Lavagno and D. Pigato}
\affiliation{Department of Applied Science and Technology, Politecnico di Torino, I-10129 Torino, Italy and \\
INFN, Sezione di Torino, I-10126 Torino, Italy}

\begin{abstract}

We investigate the possible thermodynamic instability in a warm and dense nuclear medium ($T\le 50$ MeV and $\rho_0\le\rho_B\le 3\,\rho_0$) where a phase transition from nucleonic matter to resonance-dominated $\Delta$ matter can take place. The analysis is performed by requiring the global conservation of baryon and electric charge numbers in the framework of a relativistic equation of state. Similarly to the liquid-gas phase transition, we show that the nucleon-$\Delta$ matter phase transition is characterized by both mechanical instability (fluctuations on the baryon density) and by chemical-diffusive instability (fluctuations on the charge concentration) in asymmetric nuclear matter. We then perform an investigation and a comparative study on the different nature of such instabilities and phase transitions.

\end{abstract}
\pacs{21.65.--f, 25.75.--q, 64.10.+h}
\maketitle

\section{Introduction}

One of the most interesting aspects of the experiments on heavy-ion collisions is a detailed study of the thermodynamical properties of strongly interacting nuclear matter away from the nuclear ground state. In this direction, many efforts have been focused on searching for possible phase transitions in such collisions. At low temperatures ($T\le 10$ MeV) and subnuclear densities, a liquid-gas type of phase transition was first predicted theoretically \cite{mekjian,kapusta,mishustin} and later observed experimentally in a nuclear multifragmentation phenomenon at intermediate-energy nuclear reactions \cite{pocho,xu}.

Because nuclei are made of neutrons and protons, the nuclear liquid-gas phase transition is in a binary
system where one has to deal with two independent proton and neutron chemical potentials for baryon number and electric charge  conservation. Taking into account of this important property, a very detailed study of M\"uller and Serot \cite{mullerserot} focused on the main thermodynamic properties of asymmetric nuclear matter in the framework of a relativistic mean field model.

A relevant aspect of a system with two conserved charges (baryon and isospin numbers) is that the phase transition is of second order from the viewpoint of Ehrenfest's definition. At variance with the so-called Maxwell construction for one conserved charge, the pressure is not constant in the mixed phase and therefore the incompressibility does not vanish \cite{mullerserot,prl2007}. Such feature plays a crucial role in the structure and in the possible hadron-quark phase transition in compact star objects \cite{glendenning,astro}.
Moreover, for a binary system with two phases, the binodal coexistence surface is two dimensional and the instabilities in the mixed liquid-gas phase arise from fluctuations in the proton concentration (chemical instability) and in the baryon density (mechanical instability) \cite{mullerserot,barranco,das,baran}.

Although the equation of state (EOS) at densities below the saturation nuclear matter is relatively well known due to the large amount of experimental nuclear data available, at larger densities there are many uncertainties; the strong repulsion at short distances of nuclear force makes, in fact, the compression of nuclear matter quite difficult. However, in relativistic heavy-ion collisions the baryon density can reach values of a few times the saturation nuclear density $\rho_0$ and/or high temperatures. The future CBM (compressed baryonic matter) experiment of the FAIR project at GSI Darmstadt will make it possible to create compressed baryonic matter with a high net baryon density \cite{senger}. In this direction very interesting results have been obtained at low energy at the CERN Super Proton Collider (SPS) and at low-energy scan at the BNL Relativistic Heavy Ion Collider (RHIC) \cite{alt,agga,richa}.

In regime of finite values of density and temperature, a state of high density resonance matter may be formed and the $\Delta(1232)$-isobar degrees of freedom are expected to play a central role in relativistic heavy ion collisions and in the physics of compact stars \cite{hofmann,zabrodin,xiang,chen,prc2010}. Transport model calculations and experimental results indicate that an excited state of baryonic matter is dominated by the $\Delta$-resonance at the energy from the BNL Alternating Gradient Synchrotron (AGS) to RHIC \cite{bass,mao,fachini,star-res}.
Moreover, in symmetric nuclear matter and in the framework of a non-linear Walecka model, it has been predicted that a phase transition from nucleonic matter to $\Delta$-excited nuclear matter can take place and the occurrence of this transition sensibly depends on the $\Delta$-meson coupling constants \cite{greiner87,greiner97}. Due to the presence of only one conserved "charge" (baryon number) considered in these previous investigations, the region of the phase transition develops when the incompressibility becomes negative and therefore only mechanical instabilities are present.

The information coming from experiments with heavy ions in intermediate- and high-energy collisions is that the EOS depends on the energy beam but also sensibly on the electric charge fraction $Z/A$ of the colliding nuclei, especially at not too high temperatures \cite{ditoro2006,bao}. Moreover, the study of nuclear matter with arbitrary electric charge fraction results to be important in radioactive beam experiments and in the physics of compact stars.

In this article, we study the hadronic EOS at finite temperature and density by means of a relativistic mean-field model with the inclusion $\Delta$-isobars and by requiring the Gibbs conditions on the global conservation of baryon number and net electric charge. In this context, let us observe that, for the range of temperatures and baryon densities considered in this investigation ($T\leq 50$ and  $\rho_B\le 3 \rho_0$), the contribution of strange hadron particles can be neglected in a good approximation due to their very low concentration. In fact, unlike compact stars in a $\beta$-stability regime, since weak decays cannot take place during the short lifetime of a high density system, the only possibility of producing strangeness is through associated production but, in the scenario we are discussing, this process has been shown to be very inefficient \cite{ferini,fuchs} and, therefore, the study of the possible phase transition can be limited to two conserved charges.

The main goal of this paper is to show that, for an asymmetric warm and dense nuclear medium, the possible $\Delta$-matter phase transition is characterized by mechanical and chemical-diffusive instabilities. Similarly to the liquid-gas phase transition, chemical instabilities play a crucial role in the characterization of the phase transition and can imply a very different electric charge fraction $Z/A$ in the coexisting phases during the phase transition.

The paper is organized as follows. In Sec. II, we present the relativistic hadronic equation of state.
In Sec. III, we review the most important thermodynamic proprieties of the phase transitions in a binary system, highlighting the relevant features of mechanical and diffusive instabilities. The main results are presented in Sec. IV, which is divided into two subsections: in A, we review the most important results obtained in the liquid-gas phase transition and, in B, we study the nucleon-$\Delta$ matter phase transition. Finally, in Sec. V, we summarize our conclusions.

\section{Hadronic equation of state}
\label{hadron}

The relativistic mean-field model (RMF) is widely successful used
for describing the properties of finite nuclei as well as hot and dense
nuclear matter \cite{serotwal,boguta,sharma,toki,glenmos}.

In the RMF model the Lagrangian density for nucleons can be written as
\begin{eqnarray}\label{lagrangian}
{\cal L}_{\rm N} &=&
\overline{\psi}_N\,[i\,\gamma_{\mu}\,\partial^{\mu}-(M_N-
g_{\sigma N}\,\sigma) -g_{\omega N}\,\gamma_\mu\,\omega^{\mu}\nonumber\\
&&-g_{\rho N}\,\gamma_{\mu}\,\vec{t} \cdot \vec{\rho}^{\;\mu}]\,\psi_N
+\frac{1}{2}(\partial_{\mu}\sigma\partial^{\mu}\sigma-m_{\sigma}^2\sigma^2)
-U(\sigma)\nonumber\\
&& +\frac{1}{2}\,m^2_{\omega}\,\omega_{\mu}\omega^{\mu}
+\frac{1}{4}\,c\,(g_{\omega N}^2\,\omega_\mu\omega^\mu)^2
+\frac{1}{2}\,m^2_{\rho}\,\vec{\rho}_{\mu}\cdot\vec{\rho}^{\;\mu} \nonumber\\
&&-\frac{1}{4}F_{\mu\nu}F^{\mu\nu}
-\frac{1}{4}\vec{G}_{\mu\nu}\vec{G}^{\mu\nu}\,,
\end{eqnarray}
where $M_N=939$ MeV is the nucleon vacuum mass and $\vec{t}$ is the isospin operator which acts on the nucleon.
The field strength tensors for the vector mesons are given by the usual expressions
$F_{\mu\nu}\equiv\partial_{\mu}\omega_{\nu}-\partial_{\nu}\omega_{\mu}$,
$\vec{G}_{\mu\nu}\equiv\partial_{\mu}\vec{\rho}_{\nu}-\partial_{\nu}\vec{\rho}_{\mu}$,
and $U(\sigma)$ is the nonlinear potential of $\sigma$ meson
\begin{eqnarray}
U(\sigma)=\frac{1}{3}a(g_{\sigma N}\sigma)^{3}+\frac{1}{4}b(g_{\sigma N}\sigma)^{4}\,,
\end{eqnarray}
usually introduced to achieve a reasonable compression modulus for equilibrium normal nuclear matter \cite{boguta}.
In the following, the meson-nucleon coupling constants and the other parameters ($a$, $b$, $c$) of the EOS will be fixed to the parameters set marked as TM1 of Ref. \cite{toki}.

In a regime of finite values of temperature and density, a state of high-density resonance matter may be formed and the $\Delta(1232)$-isobar degrees of freedom are expected to play a central role \cite{hofmann,zabrodin,bass,fachini}. In particular, the formation of resonances matter contributes essentially to baryon stopping, hadronic flow effects and enhanced strangeness \cite{mattiello}.

It is well known that thus far there is no relativistic quantum theory for the $\Delta$ as a spin 3/2 field without any inconsistency when imposing other fields such as the ones with electromagnetic interaction \cite{john}. Moreover, following the Rarita-Schwinger formalism, the spin 3/2 particle, described by means of a vector spinor state, has an off-shell spin 1/2 sector. To incorporate $\Delta$-isobars in the framework of effective hadron field theories, a formalism was developed to treat $\Delta$ analogously to the nucleon, taking only the on-shell $\Delta$s into account and the mass of the $\Delta$s are substituted by the effective one in the relativistic mean field approximation \cite{dejong,boguta1982}. The Lagrangian density concerning the $\Delta$-isobars can be then expressed as \cite{greiner97,boguta1982,kosov}
\begin{eqnarray}
{\mathcal L}_\Delta=\overline{\psi}_{\Delta\,\nu}\, [i\gamma_\mu
\partial^\mu -(M_\Delta-g_{\sigma\Delta}
\sigma)-g_{\omega\Delta}\gamma_\mu\omega^\mu
 ]\psi_{\Delta}^{\,\nu} \, ,
\end{eqnarray}
where $\psi_\Delta^\nu$ is the Rarita-Schwinger spinor
for the $\Delta$-isobars ($\Delta^{++}$, $\Delta^{+}$, $\Delta^0$, $\Delta^-$). Due to the uncertainty on the meson-$\Delta$ coupling constants, we limit ourselves to consider
only the coupling with the $\sigma$ and $\omega$ meson fields, more of which are explored in the literature (see Sec. IV for details)  \cite{greiner97,kosov,jin}.

In the RMF approach baryons are considered as Dirac quasiparticles
moving in classical meson fields and the field operators are
replaced by their expectation values. As a consequence, the field equations in a mean field approximation are
\begin{eqnarray}
&&(i\gamma_{\mu}\partial^{\mu}-M_N^*-
g_{\omega N}\gamma^{0}\omega-g_{\rho N}\gamma^{0}{t_{3}}\rho)\psi_N=0\,, \\
&&(i\gamma_{\mu}\partial^{\mu}-M^*_{\Delta}-
g_{\omega \Delta}\gamma^{0}\omega)\psi_{\Delta}^{\,\nu}=0\,, \\
&&m_{\sigma}^2\sigma+ a g^3_{\sigma N}{{\sigma}^2}+ b g^4_{\sigma N}{{\sigma}^3}=\sum_{i}g_{\sigma i}\rho^S_i\,, \\
&&m^2_{\omega}\omega +cg^4_{\omega N}\omega^3=\sum_{i}g_{\omega i}\rho^B_i\,, \\
&&m^2_{\rho}\rho=\sum_{i}g_{\rho i}t_{3 i}\rho^B_i\, ,
\label{eq:MFT}
\end{eqnarray}
where $\sigma=\langle\sigma\rangle$, $\omega=\langle\omega^0\rangle$ and $\rho=\langle\rho^0_3\rangle$, are the nonvanishing expectation values of meson fields, the index $i$ runs over the considered baryon particles, and the effective mass of the $i$th baryon is defined as
\begin{eqnarray}
M^*_i= M_i- g_{\sigma i}\sigma \, .  \label{eq:Meff}
\end{eqnarray}
The $\rho^B_i$ and $\rho^S_i$ are the baryon
density and the baryon scalar density, respectively. They are
given by
\begin{eqnarray}
&&\rho^{B}_i= \gamma_i \int\frac{{\rm
d}^3k}{(2\pi)^3}[n_i(k)-\overline{n}_i(k)]\,, \label{eq:rhob} \\
&&\rho^S_i= \gamma_i \int\frac{{\rm
d}^3k}{(2\pi)^3}\,\frac{M_i^*}{E_i^*}\,
[n_i(k)+\overline{n}_i(k)] \, ,  \label{eq:rhos}
\end{eqnarray}
where $\gamma_i$ is the degeneracy spin factor ($\gamma_{N}=2$ and $\gamma_{\Delta}=4$) and $n_i(k)$ and $\overline{n}_i(k)$ are the fermion particle and antiparticle
distribution functions, given by
\begin{eqnarray}
n_i(k)= \frac{1} { \exp(E_i^*(k)-\mu_i^*)/T + 1} \,, \label{eq:distribuz} \\
\overline{n}_i(k)=  \frac{1}  {\exp(E_i^*(k)+\mu_i^*)/T + 1} \, .  \label{eq:distribuz2}
\end{eqnarray}

The baryon effective energy is defined as
${E_i}^*(k)=\sqrt{k^2+{{M_i}^*}^2}$ and the effective chemical
potentials $\mu_i^*$  are given in terms of the meson fields as follows:
\begin{eqnarray}
\mu_i^*={\mu_i}-g_{\omega i}\omega - g_{\rho i}t_{3i}\rho \, ,
\label{mueff}
\end{eqnarray}
where $\mu_i$ are the thermodynamical chemical potentials,
$\mu_i=\partial\epsilon/\partial\rho_i$.

Because we are going to describe a finite temperature and density
asymmetric nuclear matter, we have to require the conservation of two "charges": baryon
number ($B$) and electric charge ($C$) (as already remarked, we neglect the contribution of strange hadrons,
because a tiny amount of strangeness can be produced in the range of temperature and density explored in this study).
As a consequence, the system is described by two independent chemical potentials: $\mu_B$ and
$\mu_C$, the baryon and the electric charge chemical potential, respectively. Therefore,
the chemical potential of particle of index $i$ can be written as
\begin{equation}
\mu_i=b_i\, \mu_B+c_i\,\mu_C \, , \label{mu}
\end{equation}
where $b_i$ and $c_i$ are, respectively, the baryon and the
electric charge quantum numbers of the $i$th hadron.

The thermodynamical quantities can be obtained from the baryon grand
potential $\Omega_B$ in the standard way. More explicitly, the
baryon pressure $P_B=-\Omega_B/V$ and the energy density can be
written as
\begin{eqnarray}
P_B&=&\frac{1}{3}\sum_i \,\gamma_i\,\int \frac{{\rm
d}^3k}{(2\pi)^3}
\;\frac{k^2}{E_{i}^*(k)}\; [n_i(k)+\overline{n}_i(k)] \nonumber \\
&-&\frac{1}{2}\,m_\sigma^2\,\sigma^2 - U(\sigma)+
\frac{1}{2}\,m_\omega^2\,\omega^2+\frac{1}{4}\,c\,(g_{\omega
N}\,\omega)^4  \nonumber \\
&+&\!\!\frac{1}{2}\,m_{\rho}^2\,\rho^2 ,\\
\epsilon_B&=&\sum_i \,\gamma_i\,\int \frac{{\rm
d}^3k}{(2\pi)^3}\;E_{i}^*(k)\; [n_i(k)+\overline{n}_i(k)]\nonumber \\
&+&\frac{1}{2}\,m_\sigma^2\,\sigma^2
+U(\sigma)+\frac{1}{2}\,m_\omega^2\,\omega^2+\frac{3}{4}\,c\,(g_{\omega
N}\,\omega)^4 \nonumber\\
&+&\!\!\frac{1}{2}\,m_{\rho}^2
\,\rho^2 \,  .
\end{eqnarray}

Let us observe that the contribution of the lightest non-strange mesons (pions) may not be negligible in regime of temperature and density achieved during the possible $\Delta$-matter phase transition. Following Ref. \cite{muller}, from a phenomenological point of view, we can take into account the pion degrees of freedom by adding their one-body contribution to the thermodynamical potential, that is, the contribution of an ideal Bose gas with an effective pion chemical potential $\mu_\pi^*$, depending self-consistently from the meson fields. The value of $\mu_\pi^*$ is obtained from the "bare" one $\mu_\pi$, given from Eq. (\ref{mu}), and subsequently expressed in terms of the corresponding effective baryon chemical potentials, respecting the strong interaction. More explicitly, from Eq. (\ref{mu}),
$\mu_{\pi^+}=\mu_C\equiv\mu_p-\mu_n$ and the corresponding effective chemical potential can be written as
\begin{eqnarray}
\mu_{\pi^+}^*&\equiv&\mu_p^*-\mu_n^*=\mu_p-\mu_n-g_{\rho N}\,\rho \, , \label{mueff_m1}
\end{eqnarray}
where the last equivalence follows from Eq. (\ref{mueff}). As a consequence, $\mu_{\pi^-}^*= - \mu_{\pi^+}^*$ and $\mu_{\pi^0}^*\equiv\mu_{\pi^0}=0$.

This assumption can be seen somehow in analogy with the hadron resonance gas within the excluded-volume approximation. There the hadronic system is still regarded as an ideal gas but in the volume reduced by the volume occupied by constituents (usually assumed as a phenomenological model parameter), here we have a (quasifree) pion gas but with an effective chemical potential that contains the self-consistent interaction of the meson fields.

Of course, this naive phenomenological approach cannot incorporate the very complex $\pi N \Delta$ interaction at finite temperature and baryon density and a more realistic chiral symmetric model should be implemented. On the other hand, as we will see in Sec. IV-B, such an effective nuclear EOS has the noticeable advantage of simplifying the not trivial numerical analysis involved in seeking of thermodynamic instabilities and in the construction of  the mixed phase. Due to this fact, it would be prudent to see the results of this preliminary study in a perspective of academic interest.

Finally, the total pressure and energy density are given by the baryon (B) and pion (M) contribution:  $\epsilon= \epsilon_B +\epsilon_M$ and $P=P_B +P_M$.

\section{Phase transitions and stability conditions}

As already stated, we are dealing with the study of a multi-component system at finite temperature and density with two conserved charges: baryon number and electric charge. For such a system, the Helmholtz free energy density $F$ can be written as
\begin{equation}
F(T,\rho_B,\rho_C)= -P(T,\mu_B,\mu_C) +\mu_B\rho_B + \mu_C\rho_C \, ,
\end{equation}
with
\begin{equation}
\mu_B=\left (\frac{\partial F}{\partial\rho_B}\right)_{T,\rho_C} \, , \ \ \ \mu_C=\left ( \frac{\partial F}{\partial\rho_C}\right)_{T,\rho_B} \, .
\end{equation}
In a system with $N$ different particles, the particle chemical potentials are expressed as the linear combination of the two independent chemical potentials $\mu_B$ and $\mu_C$ and, as a consequence,  $\sum_{i=1}^N \mu_i\rho_i=\mu_B\rho_B + \mu_C\rho_C$. Therefore, the number of particles may change during a process and, at variance of density and temperature, different particle degrees of freedom may be relevant in the description of the system (for example, at low temperature and density, we have protons and neutrons only, while at higher temperature and density other kind of particles, such as $\Delta$-isobars, can appear). What it is actually relevant for the thermodynamical description under consideration are only the two conserved charges and not the number of different particles constituent the system.

In general, a system can exist in a number of different phases, each of which exhibit quite different macroscopic behavior. The single phase that is realized for a given set of independent variables is the one with the lowest free energy. In a system with two conserved charges, it is possible to have $N_{\rm max}=4$ phase coexistence regions in thermodynamical equilibrium  \cite{landau,reichl}, even if we have found no evidence for the existence of more than two phases in the regime investigated in this paper. Therefore,
assuming the presence of two phases (denoted as $I$ and $II$, respectively), the system is stable against the separation in two phases if the free energy of a single phase is lower than the free energy in all two phases configuration. The phase coexistence is given by the Gibbs conditions
\begin{eqnarray}
&&\mu_B^{I} = \mu_B^{II} \, , \ \ \ \ \ \ \ \ \ \mu_C^{I} = \mu_C^{II}
\, , \\
&&P^I (T,\mu_B,\mu_C)=P^{II} (T,\mu_B,\mu_C) \, .
\end{eqnarray}
Therefore, at a given baryon density $\rho_B$ and at a given net
electric charge density $\rho_C=y\, \rho_B$ (with $y=Z/A$), the chemical potentials $\mu_B$ are $\mu_C$ are univocally determined by the
following equations
\begin{eqnarray}
&&\!\!\!\!\!\!\!\!\rho_B=(1-\chi)\,\rho_B^I(T,\mu_B,\mu_C) +\chi \,\rho_B^{II}(T,\mu_B,\mu_C) \, ,\label{rhobchi}\\
&&\!\!\!\!\!\!\!\!\rho_C=(1-\chi)\,\rho_C^I(T,\mu_B,\mu_C) +\chi \,\rho_C^{II}(T,\mu_B,\mu_C) \, ,
\label{rhocchi}\end{eqnarray}
where $\rho_B^{I(II)}$ and $\rho_C^{I(II)}$ are, respectively, the baryon and electric charge densities in the low density ($I$) and in the higher density ($II$) phase and $\chi$ is the volume fraction of the phase $II$ in the mixed phase ($0\le\chi\le 1$).

An important feature of this conditions is that, unlike the case of a single conserved charge, the pressure in the mixed phase is not constant and, although the total $\rho_B$ and $\rho_C$ are fixed, baryon and charge densities can differ in the two phases, according to Eq.s (\ref{rhobchi}) and (\ref{rhocchi}).

For such a system in thermal equilibrium, the possible phase transition can be characterized by mechanical (fluctuations in the baryon density) and chemical instabilities (fluctuations in the electric charge density).
As usual the condition of the mechanical stability implies
\begin{eqnarray}
\rho_B \left(\frac{\partial P}{\partial \rho_B}\right)_{T,\,\rho_C} >0   \, .  \label{InstabMecc}
\end{eqnarray}
By introducing the notation $\mu_{i,j}=(\partial\mu_i/\partial\rho_j)_{T,P}$ (with $i,j=B,C$), the chemical stability can be expressed with the following conditions \cite{reichl}
\begin{eqnarray}
\mu_{B,B} >0 \, , \ \ \ \mu_{C,C}>0 \, , \ \ \
\begin{vmatrix}
\,\mu_{B,B} & \mu_{B,C} \\
\,\mu_{C,B} & \mu_{C,C}
\end{vmatrix}
>0  \,  . \label{InstabChim}
\end{eqnarray}
In addition to the above conditions, for a process at constant $P$ and $T$, it is always satisfied that
\begin{eqnarray}
&&\rho_B\, \mu_{B,B}+\rho_C \, \mu_{C,B}=0\, , \\
&&\rho_B\, \mu_{B,C}+\rho_C \, \mu_{C,C}=0\, . \label{diff1}
\end{eqnarray}

Whenever the above stability conditions are not respected, the system becomes unstable and the phase transition take place.  The coexistence line of a system with one conserved charge becomes in this case a two dimensional surface in $(T,P,y)$ space, enclosing the region where mechanical and diffusive instabilities occur.

\section{Results and discussion}\label{result}

As mentioned in the Introduction, our main goal is to study the instability regions related to the formation of the $\Delta$-isobars at finite temperature and baryon density. Because such instabilities may have several analogies with the liquid-gas phase transition, it is instructive, first, to briefly review the main properties of this nuclear phase transition in the framework of our EOS and to test the numerical procedure that will be applied at higher temperatures and densities.

\subsection{Liquid-gas phase transition}

In a regime of low temperature and baryon density, relevant in the liquid-gas phase transition, only proton and neutron degrees of freedom take place.
In this simple case, for example, Eq. (\ref{diff1}) can be written as
\begin{equation}
y\, \left(\frac{\partial \mu_p}{\partial y}\right)_{T,P}+(1-y)\, \left(\frac{\partial \mu_n}{\partial y}\right)_{T,P}=0 \, ,\label{diff2}
\end{equation}
where $y=\rho_p/\rho_B$. Because we are working with a proton fraction $0< y\le 0.5$, the chemical stability conditions (\ref{InstabChim}) are therefore satisfied if
\begin{equation}
\left(\frac{\partial \mu_p}{\partial y}\right)_{T,P}>0  \ \ {\rm or } \ \ \left(\frac{\partial \mu_n}{\partial y}\right)_{T,P}<0 \,
\label{inst_lg}
\end{equation}
[due the validity of Eq. (\ref{diff2}), the first above condition implies the second one and vice versa].

\begin{figure}
\begin{center}
\resizebox{0.45\textwidth}{!}{%
\includegraphics{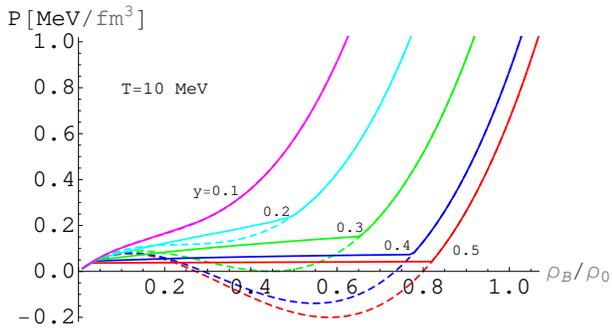}
} \caption{(Color online) Pressure as a function of baryon density for various values of the proton fraction. The continuous (dashed) lines correspond to the solution obtained with (without) the Gibbs construction.} \label{fig:P_rhob_lg}
\end{center}
\end{figure}

As already observed, in presence of two conserved charges the liquid-gas phase transition can be characterized by mechanical and chemical instabilities \cite{mullerserot}. In order to better put this feature in focus, we report
in Fig. \ref{fig:P_rhob_lg} the pressure as a function of baryon density for various values of the electric charge fraction $y$ at fixed temperature $T=10$ MeV. The continuous lines correspond to the solution obtained with the Gibbs construction, whereas the dashed lines are without correction.
For a proton fraction $y>0.2$ a mechanical instability is present, whereas for $y<0.2$ the system becomes unstable only under chemical-diffusive instability.

\begin{figure}
\begin{center}
\resizebox{0.45\textwidth}{!}{%
\includegraphics{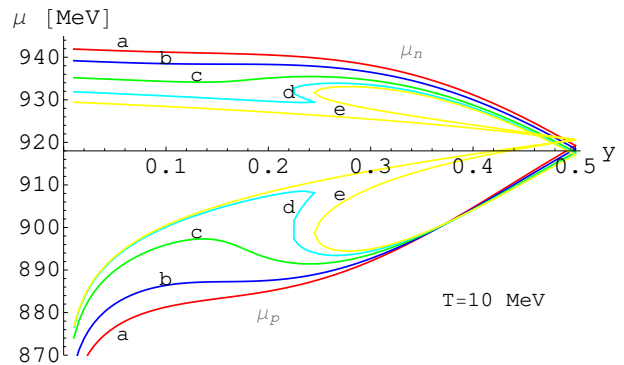}
} \caption{(Color online) Proton and neutron chemical potential as a function of the proton fraction $y$ for various isobars ($P$=0.25, 0.20, 0.15, 0.10, 0.075 MeV/fm$^3$) (lines $a$ to $e$) at $T$=10 MeV.
} \label{fig:mu-za}
\end{center}
\end{figure}

The presence of chemical unstable regions are much more evident in Fig. \ref{fig:mu-za}, where we show the proton and neutron chemical potentials for various isobars at constant temperature, as a function of the proton asymmetry.
Below $P=0.25$ MeV/fm$^3$, the system becomes un\-stable because of the presence of regions of negative (positive) slope for $\mu_p$ ($\mu_n$).
%

%
%

\begin{figure}
\begin{center}
\resizebox{0.45\textwidth}{!}{%
\includegraphics{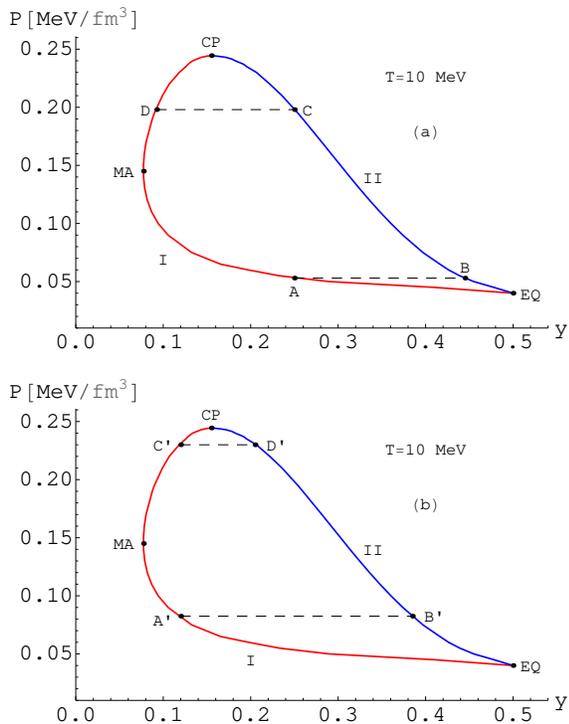}
} \caption{(Color online) Binodal section at $T=10$ MeV, with in evidence the critical point (CP), the point of maximum asymmetry (MA) and the point of equal equilibrium (EQ). In the upper and lower panels  are reported the evolution of the mixed phase for two different system configurations (see the text for details).} \label{fig:P_ZA}
\end{center}
\end{figure}

In order to study the phase coexistence of the system, in Fig. \ref{fig:P_ZA}, we show the binodal section as a function of the proton asymmetry $y$ at $T=10$ MeV.
Following the same notation of Ref. \cite{mullerserot}, the binodal surface is divided into two branches by a critical point (CP) and a point of equal equilibrium (EQ) at $y=0.5$, where protons and neutrons have the same concentration. The left branch of the diagram represents the initial phase configuration of the system at lower density (gas phase, I) and the second branch, at higher density, corresponds to the final phase configuration (liquid phase, II).

The binodal surface encloses the area where the system undergoes to the phase transition. The mixed-phase region extends up to small values of the proton asymmetry, whereas the mechanical instability region ends around $y\simeq 0.2$.

During the isothermal compression, the system evolves through configuration at constant $y$ and meets the first branch in a point $A$. At this point the system becomes unstable and an infinitesimal phase in $B$ appears at the same temperature and pressure of $A$. In this context let us remember that, although the proton asymmetry is globally conserved, this is not true for the single phase. In particular, for an asymmetric nuclear system it is energetically favorable to separate it into a liquid phase (less asymmetric) and a gas phase (more asymmetric) rather than into two phases with equal proton fraction.

If point $A$ has a value of $y_A$ greater than the corresponding values $y_{\rm CP}$ of the CP (as in the upper panel of Fig. \ref{fig:P_ZA}), the system ends the phase transition in the liquid phase (in the point $C$).
On the other hand, as already observed in Ref. \cite{mullerserot}, if the system has been prepared in a very asymmetric configuration with $y_{A^{'}}< y_{\rm CP}$ (lower panel of Fig. \ref{fig:P_ZA}), it undergoes to a retrograde phase transition. A second liquid phase in $B^{'}$ is formed but, after reaching a point of maximum volume fraction $\chi_{\rm max}<1$, the system returns to its initial gas phase at point $C^{'}$. Note that this kind of phase transition is possible only for a multi-component system and in this case, is purely diffusive.

\begin{figure}[tb]
\begin{center}
\resizebox{0.45\textwidth}{!}{%
\includegraphics{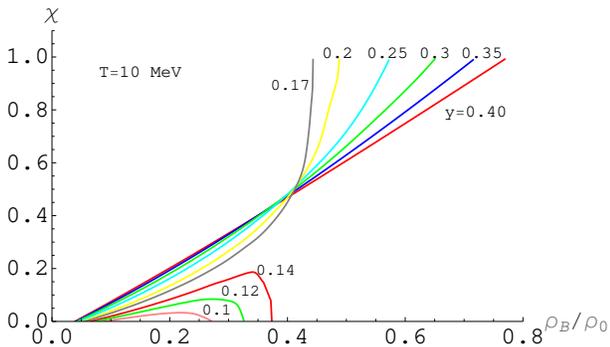}
} \caption{(Color online) Evolution of the volume fraction $\chi$ of the second phase as a function of the baryon density for a system with different values of $y$ at $T=10$ MeV.} \label{fig:chi_rhob}
\end{center}
\end{figure}
To better characterize the evolution of the two phases, in Fig. \ref{fig:chi_rhob}, the volume fraction $\chi$ of the second phase during the phase transition is shown. By increasing the asymmetry parameter of the system under consideration (at lower values of $y$), the maximum density achieved during the mixed phase decreases, until the system undergoes a retrograde phase transition (for $y< 0.15$).


\subsection{$\Delta$-matter phase transition}


By increasing the temperature and the baryon density during the high energy heavy ion collisions ($T\approx 50$ MeV and $\rho_0\le\rho_B\le 3\,\rho_0$), a multi-particle system with $\Delta$-isobar and pion degrees of freedom may take place.

\begin{figure}
\begin{center}
\resizebox{0.45\textwidth}{!}{%
\includegraphics{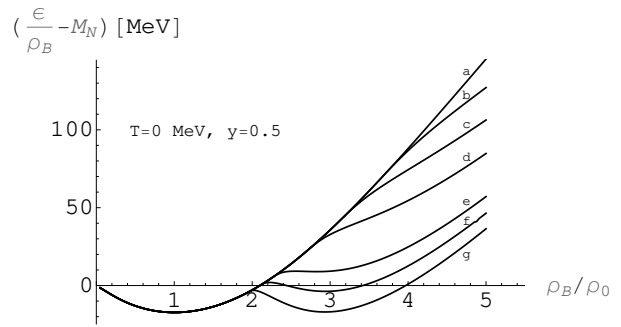}
} \caption{The energy per baryon versus baryon density at zero temperature and $y=0.5$ with (a) no  $\Delta$; (b) $x_{\sigma \Delta}=1.10$; (c) $x_{\sigma \Delta}=1.15$; (d) $x_{\sigma \Delta}=1.20$; (e) $x_{\sigma \Delta}=1.27$; (f) $x_{\sigma \Delta}=1.30$; (g) $x_{\sigma \Delta}=1.33$.}
\label{fig:Enuc}
\end{center}
\end{figure}

To better understand the relevance of $\Delta$-isobar and the dependence of the EOS on the meson-$\Delta$ coupling constants ($x_{\sigma\Delta}=g_{\sigma\Delta}/g_{\sigma N}$, $x_{\omega\Delta}=g_{\omega\Delta}/g_{\omega N}$), in Fig. \ref{fig:Enuc}, we report the energy per baryon as a function of the baryon density at zero temperature and $y=0.5$ for different values of $x_{\sigma \Delta}$ and $x_{\omega\Delta}=1$. Let note that by increasing the value of $x_{\sigma \Delta}$, a second minimum on the energy per baryon appears.

Following Ref. \cite{kosov}, in setting  $x_{\sigma\Delta}$ and  $x_{\omega\Delta}$, we have to require that (i) the second minimum of the energy per baryon lies above the saturation energy of normal nuclear matter, i.e., in the mixed $\Delta$-nucleon matter only a metastable state can occur; (ii) there are no $\Delta$-isobars present at the saturation density; and (iii) the scalar field is more (equal) attractive and the vector potential is less (equal) repulsive for $\Delta$s than for nucleons, in accordance with QCD finite-density calculations \cite{jin}.
In this context, it is proper to remember that QCD sum-rule predictions for the scalar self-energy are sensitive to the unknown density dependence of four-quark condensates and, due to this, there is no certainly reliable information about the coupling constant of the $\Delta$ isobars with scalar mesons.

Of course, the choice of couplings that satisfy the above conditions is not unique but exists a finite range of possible values (represented as a triangle region in the plane $x_{\sigma\Delta}$-$x_{\omega\Delta}$) which depends on the particular EOS under consideration \cite{kosov}. Without loss of generality, in the following we can limit our investigation to move only in a side of such a triangle region by fixing $x_{\omega\Delta}=1$ and varying $x_{\sigma\Delta}$ from unity to a maximum value compatible with the aforementioned conditions. As can be observed in Fig. \ref{fig:Enuc}, for the TM1 parameter set, such a maximum value corresponds to $x_{\sigma\Delta}^{\rm max}=1.33$, while the value $x_{\sigma\Delta}^{\rm II}=1.27$ corresponds to the appearance of the second minimum on the energy per baryon with the formation of a metastable state. Analog behaviors can be obtained with other EOS parameters set (see, for example, Ref. \cite{prc2010} for more details).

\begin{figure}
\begin{center}
\resizebox{0.45\textwidth}{!}{%
\includegraphics{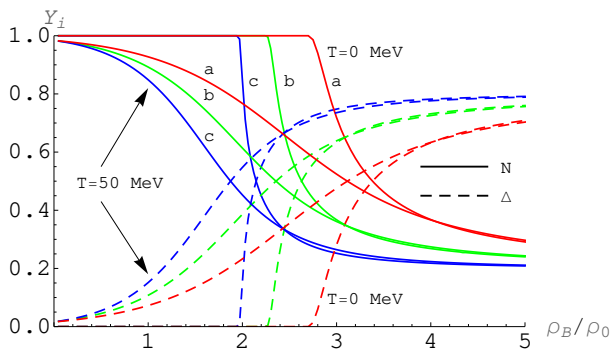}
} \caption{(Color online) The relative nucleons (solid lines) and $\Delta$ (dashed lines) densities as a function of the baryon density for different values of temperature with (a) $x_{\sigma \Delta }=1.2$; (b) $x_{\sigma \Delta }=1.27$; (c) $x_{\sigma \Delta }=1.33$.} \label{fig:Enuc2}
\end{center}
\end{figure}

In Fig. \ref{fig:Enuc2}, we show in symmetric nuclear matter the relative nucleon (solid lines) and the $\Delta$-isobar (dashed lines) density fraction ($Y_i=\rho_i/\rho_B$) versus the baryon density at $T=0$ and $T=50$ MeV, for different values of  $x_{\sigma \Delta}$. We observe that $\Delta$-matter becomes dominant with respect to the nucleon concentration at high baryon density and such effect is significantly anticipated by increasing the temperature \footnote{Let us remark that the range of baryon density reported in Figs. 5 and 6 has been chosen in order to better show the effects of different $x_{\sigma\Delta}$ couplings on the formation of $\Delta$ isobars at high baryon density. As we will see, the presence thermodynamic instabilities will be relevant at lower values of baryon density ($\rho_B\le 3\,\rho_0$).}.

In analogy with the liquid-gas case, we are going to investigate the existence of a possible phase transition in the nuclear medium by studying the presence of instabilities (mechanical and/or chemical) in the system.

As already observed, during a phase transition with two conserved charges, the electric charge fraction $y=\rho_C/\rho_B$ is not locally conserved in the single phase but only globally conserved. Therefore, during the compression of the system, the appearance of particles with negative electric charge (such as $\Delta^-$) could, in principle, shift the diffusive instability region to negative values of $y$, even if the system is prepared with a positive $y$. Such a feature has no counterpart in the liquid-gas phase transition and, as we will see, it turns out to be very relevant in order to properly determine the instability region through the binodal phase diagram.

Taking into account that Eq. (\ref{diff2}) becomes in this case
\begin{equation}
\left(\frac{\partial \mu_B}{\partial y}\right)_{T,P}+y\, \left(\frac{\partial \mu_C}{\partial y}\right)_{T,P}=0 \, ,\label{diff_delta}
\end{equation}
the chemical stability condition is satisfied if
\begin{equation}
\!\!\left(\frac{\partial \mu_C}{\partial y}\right)_{T,P}>0 \ \ {\rm or } \ \ \left\{
\begin{array}{rl}
\displaystyle
\left(\frac{\partial \mu_B}{\partial y}\right)_{T,P}<0\,, & \mbox{if } y>0 \, , \\
\\
\displaystyle
\left(\frac{\partial \mu_B}{\partial y}\right)_{T,P}>0\,, & \mbox{if } y<0 \, .
\end{array}
\right.
\end{equation}

It is relevant to observe that for the value $x_{\sigma\Delta}=1$, we do not find any mechanical or diffusive instability. Contrariwise, by increasing the $x_{\sigma \Delta}$ coupling ratio, mechanical and chemical instabilities take place. In particular, in the range $1< x_{\sigma \Delta}\le 1.1$, instabilities are restricted to very low values of temperature and electric charge fraction, but for $x_{\sigma \Delta}>1.1$, such instabilities start to be much more relevant and extend to higher values of $T$ and $y$.
\begin{figure}[htb]
\begin{center}
\resizebox{0.45\textwidth}{!}{%
\includegraphics{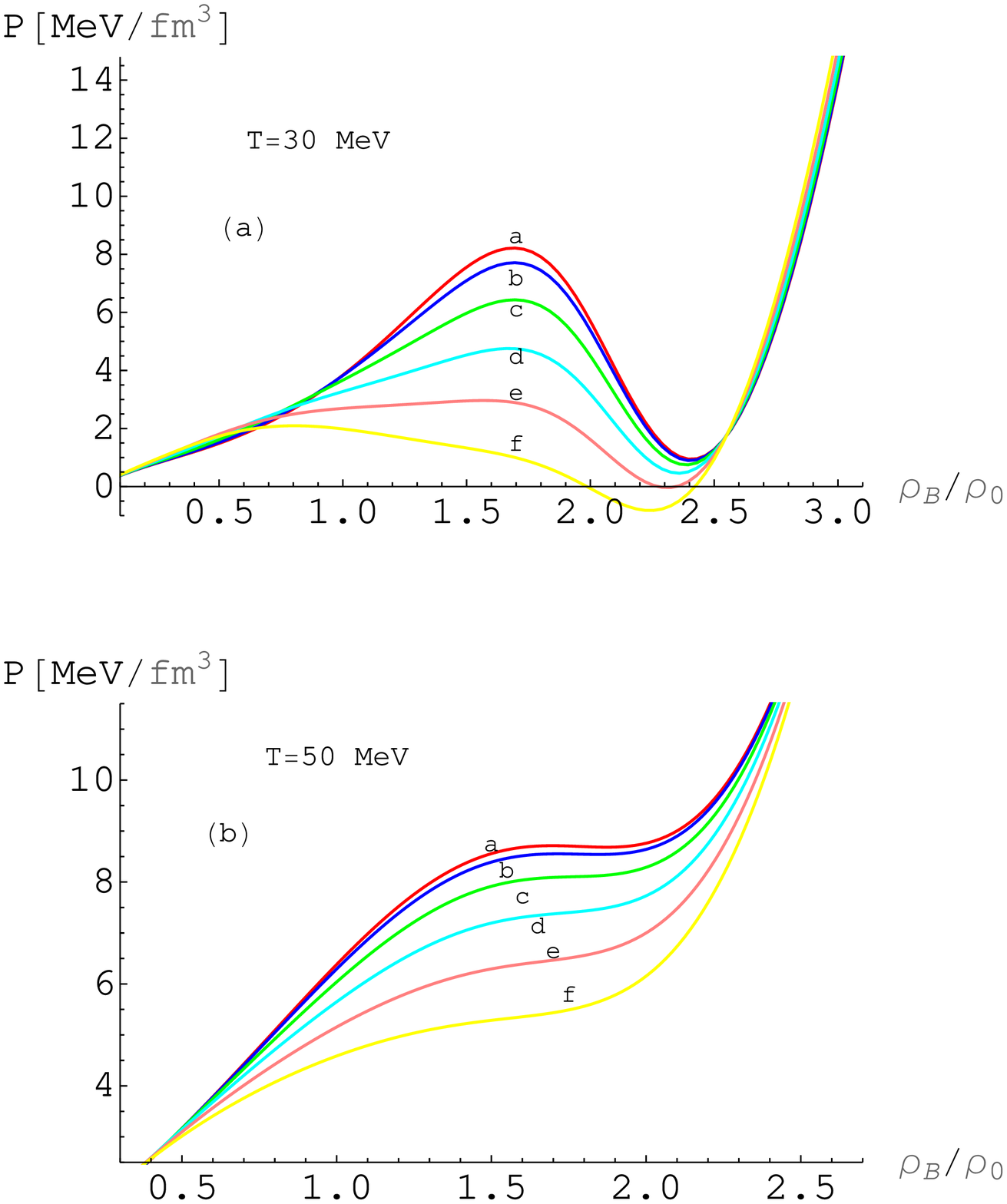}
} \caption{(Color online) Pressure as a function of baryon density for $T=30$ MeV (upper panel) and $T=50$ MeV (lower panel) with $x_{\sigma \Delta}=1.3$. Labels a to f correspond to $y=0.5,0.4,0.3,0.2,0.1,0$, respectively.} \label{fig:Prhob1}
\end{center}
\end{figure}

\begin{figure}[tb]
\begin{center}
\resizebox{0.45\textwidth}{!}{%
\includegraphics{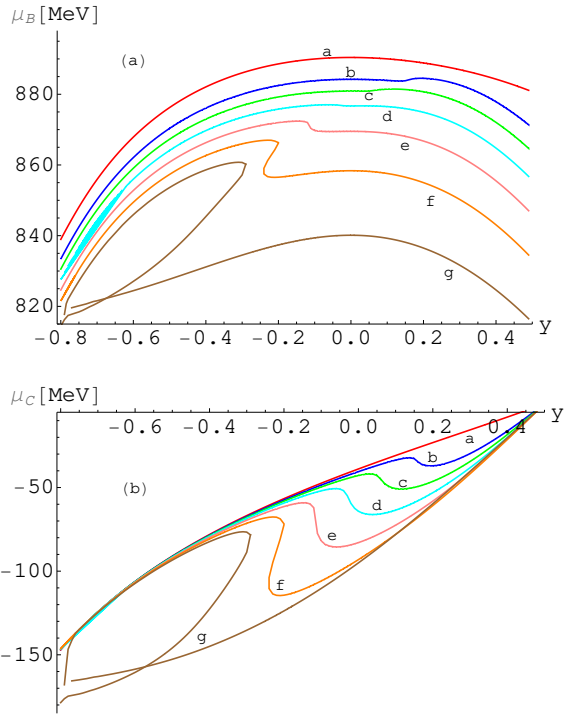}
} \caption{(Color online) Baryon (upper panel) and electric charge (lower panel) chemical potential isobars as a function of $y$ at $T=50$ MeV and $x_{\sigma \Delta}=1.3$. The curves labeled $a$ through $g$ have pressure $P$=9, 7, 6, 5, 4, 3, 2 MeV/fm$^3$, respectively.} \label{fig:muiZA}
\end{center}
\end{figure}

To better clarify this aspect, we report in Fig. \ref{fig:Prhob1}, the pressure as a function of the baryon density at $T=30$ MeV (upper panel) and $T=50$ MeV (lower panel), for different values of  $y$ and $x_{\sigma \Delta}=1.3$. In this context it is interesting to observe that at $T=50$ MeV and below $y=0.3$, the system becomes mechanically stable, but, in a similar manner to the liquid-gas case, is chemically unstable. This important feature can be better observed in the Fig. \ref{fig:muiZA}, where we report the baryon and electric charge chemical potential isobars as a function of  $y$, at fixed temperature $T=50$ MeV and $x_{\sigma \Delta}=1.3$.

\begin{figure}
\begin{center}
\resizebox{0.45\textwidth}{!}{%
\includegraphics{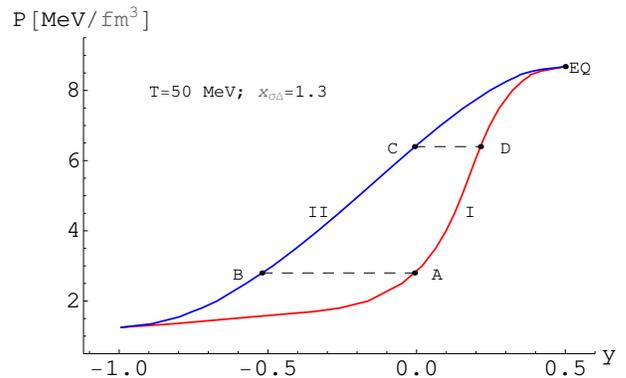}
} \caption{(Color online) Binodal section at $T=50$ MeV and $x_{\sigma \Delta}=1.3$.} \label{fig:PZA}
\end{center}
\end{figure}

From the analysis of the above chemical potential isobars, we are able to construct the binodal surface relative to the nucleon-$\Delta$ matter phase transition. In Fig. \ref{fig:PZA}, we show the binodal section at $T=50$ MeV and $x_{\sigma \Delta}=1.3$.

The right branch (at lower density) corresponds to the initial phase (I), where the dominant component of the system is given by nucleons. The left branch (II) is related to the final phase at higher densities, where the system is composed primarily by $\Delta$-isobar degrees of freedom ($\Delta$-dominant phase).  In the presence of $\Delta$ isobars, the phase coexistence region results differ substantially from what was obtained in the liquid-gas case, in particular it extends up to regions of negative electric charge fraction and the mixed phase region ends in a point of maximum asymmetry with $y=-1$ (corresponding to a system with almost all $\Delta^-$ particles, with the contribution from antiparticles and pions almost negligible in this regime).

Repeating the reasoning made for the liquid-gas phase transition, we analyze the phase evolution of the system during the isothermal compression from an arbitrary initial point $A$, indicated in Fig. \ref{fig:PZA}. In this point, the system becomes unstable and starts to be energetically favorable the separation into two phases; therefore, an infinitesimal $\Delta$-dominant phase appears in $B$, at the same temperature and pressure. Let us observe that, although in $B$ the electric charge fraction is substantially negative, the relative $\Delta^-$ abundance must be weighed on the low volume fraction occupied by phase II near point $B$ ($\chi\approx 0$). During the phase transition ($0<\chi<1$), each phase evolves towards a configuration with increasing $y$, in contrast to the liquid-gas case, where each phase evolves through a configuration with a decreasing value of $y$ (with the exception of the gas phase after the maximum asymmetry point).

\begin{figure}
\begin{center}
\resizebox{0.45\textwidth}{!}{%
\includegraphics{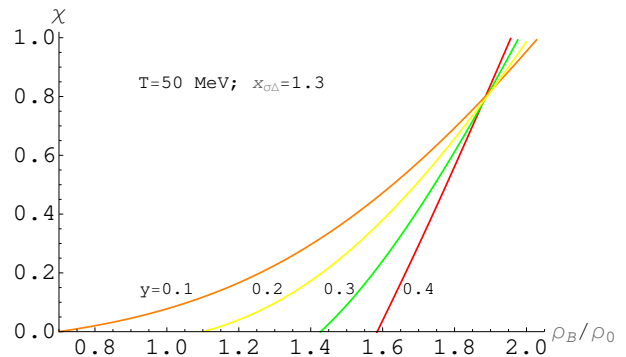}
} \caption{(Color online) Volume fraction of the $\Delta$-matter phase as a function of the baryon density for a system with different values of $y$.} \label{fig:chirhobDelta}
\end{center}
\end{figure}

In order to better understand the evolution of the two phases in the mixed phase, we report in Fig. \ref{fig:chirhobDelta} the volume fraction $\chi$ of the $\Delta$-matter phase as a function of the baryon density. Unlike in the liquid-gas case, by decreasing the electric charge fraction $y$ of the system under consideration, the mixed phase involves a greater region of baryon density and extends below the nuclear saturation density.

\begin{figure}
\begin{center}
\resizebox{0.45\textwidth}{!}{%
\includegraphics{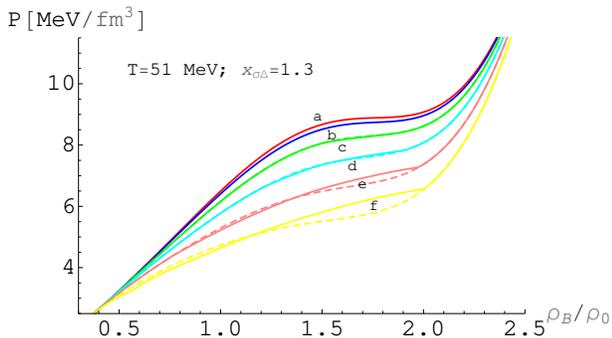}
} \caption{(Color online) Pressure as a function of baryon density at different values of $y$, from $y=0.5$ (label a) to $y=0$ (label f). The continuous (dashed) lines correspond to the solution obtained with (without) the Gibbs construction.} \label{fig:PZA2}
\end{center}
\end{figure}

In the previous example we considered a fixed value of temperature. The maximum temperature at which the system becomes mechanically stable depends on the particular value of the electric charge fraction. For example, at $y=0.3$, it is about $T_{\rm max}=49.5$ MeV and at $y=0.5$, it is about $T_{\rm max}=50.6$ MeV. Furthermore, when $y=0.5$, the end of the mechanical instability region, obviously, corresponds to the end of the mixed-phase region. This is no longer true in the presence of two conserved charges. Due to the presence of a diffusive instability region, the mixed phase can extend to slightly higher temperatures with respect to the maximum temperature achieved in the symmetric case. Although this feature involves small differences in temperature, is interesting from a conceptual point of view to investigate this aspect in more detail.
Toward this purpose, in Fig. \ref{fig:PZA2}, we show the pressure as a function of the baryon density and the Gibbs construction (continuous lines) for various values of the electric charge fraction, $x_{\sigma \Delta}=1.3$ and $T=51$ MeV (dashed lines are without Gibbs construction). In this case the system is always mechanically stable, while it is unstable for the presence of the chemical-diffusive instability up to $y=0.35$.

\begin{figure}
\begin{center}
\resizebox{0.45\textwidth}{!}{%
\includegraphics{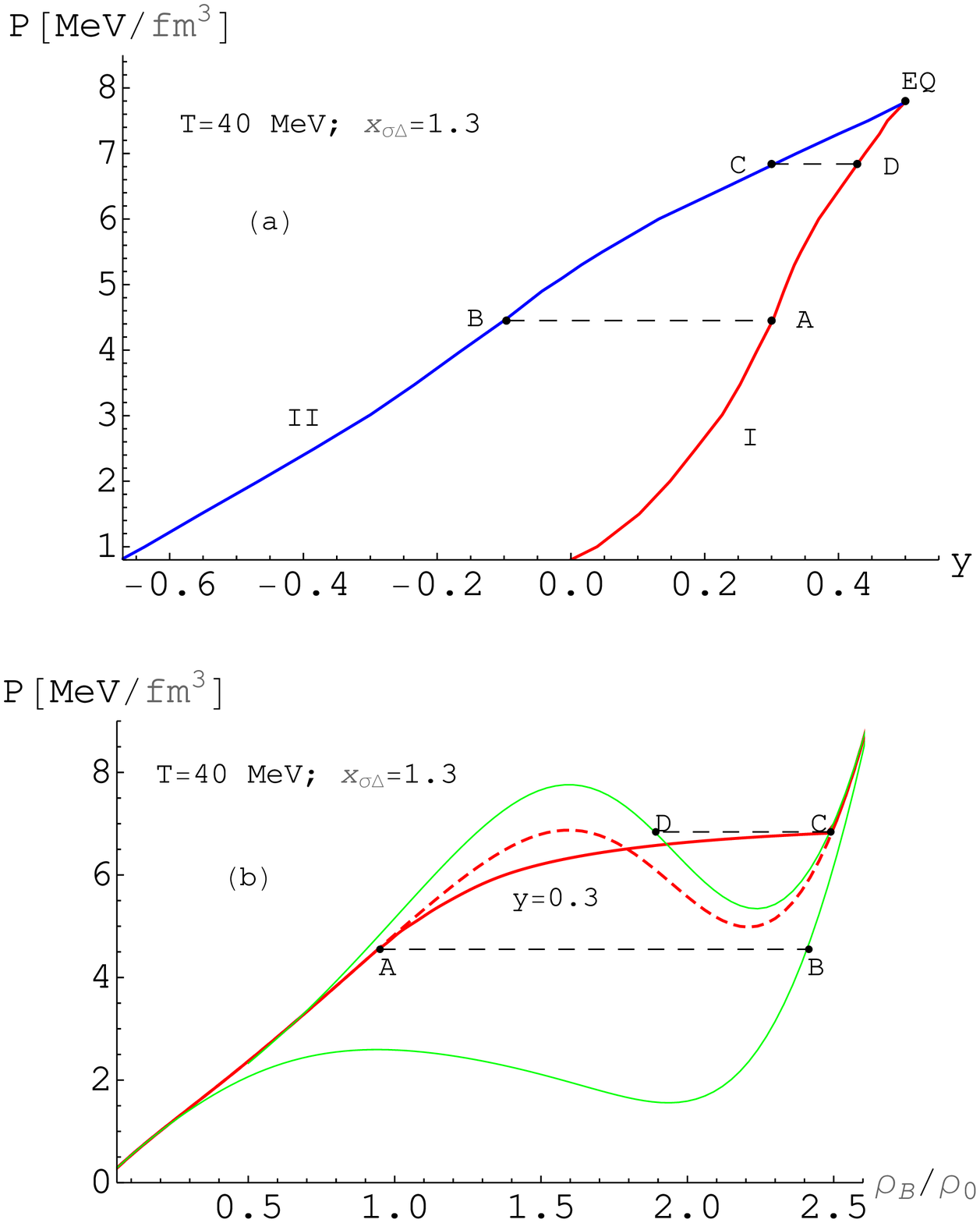}
} \caption{(Color online) Binodal section at $T$= 40 MeV and $x_{\sigma\Delta}=1.3$, with the point of equal equilibrium in evidence. (Lower panel) The corresponding isothermal curves, with the Gibbs construction (curve from the point A to C) at $y=0.3$ and the isotherms of points B and D shown.} \label{fig:T40}
\end{center}
\end{figure}

At lower temperatures, the mixed-phase region becomes more relevant at higher values of $y$. This feature can be seen in Fig. \ref{fig:T40} where the binodal section (upper panel) and the isothermal pressure as a function of the baryon density (lower panel) is reported at $T=40$ MeV and $x_{\sigma \Delta}=1.3$. The Gibbs construction corresponds to the curve from A to C; the isothermal curves in B and D (with $y_B\neq y_D$) are also reported.
In this case, we assume that the system is initially prepared in the low-density (nucleonic) phase with $y=0.3$, corresponding to point A. During the compression each phase evolves following the corresponding curve up to points C and D, where the system leaves the instability region in the $\Delta$-matter phase.
%

%

\begin{figure}
\begin{center}
\resizebox{0.45\textwidth}{!}{%
\includegraphics{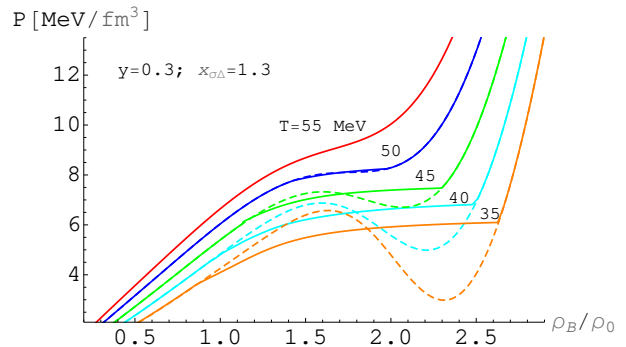}
} \caption{(Color online) Isotherms at constant $y=0.3$ and $x_{\sigma \Delta}=1.3$, for various values of temperatures. The solid (dashed) lines represent the EOS obtained with (without) Gibbs construction.} \label{fig:multiT}
\end{center}
\end{figure}

In Fig. \ref{fig:multiT}, we show the Gibbs construction (continuous lines) to the EOS at $y=0.3$, $x_{\sigma \Delta}=1.3$ and for different temperatures. By decreasing the temperature, the instability region extends over a wide range of baryon density. In particular, below $T=40$ MeV, the phase transition starts slightly below the nuclear saturation density.
\begin{figure}
\begin{center}
\resizebox{0.45\textwidth}{!}{%
\includegraphics{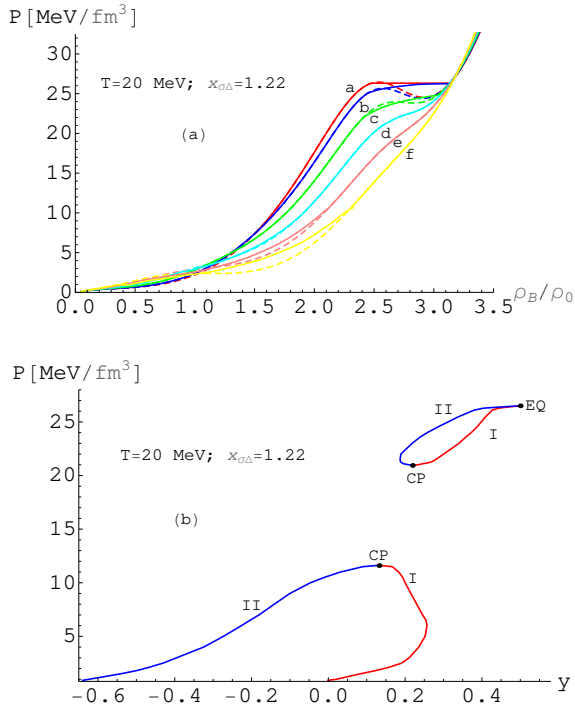}
} \caption{(Color online) (Upper panel) Pressure as a function of the baryon density for different values of $y$ from $y=0.5$ (label a) to $y=0$ (label f), with the instability regions (Gibbs construction in the continuous lines) in evidence. (Lower panel) The binodal diagram, with the point of equal equilibrium and the two critical points in evidence. In both instability sectors a region of retrograde phase transition is present.} \label{fig:PZAxsd122}
\end{center}
\end{figure}

As already observed, the mixed phase structure results strongly affected not only by the temperature, but also by the particular choice of the $x_{\sigma \Delta}$ coupling. In fact, by decreasing the $\sigma$-$\Delta$ coupling constant, the mixed phase region shifts to lower temperatures. To better clarify this aspect,
we study the phase transition for $x_{\sigma \Delta}=1.22$ and $T=20$ MeV (at $T=50$ MeV, the system results to be mechanically and chemically stable).

In Fig. \ref{fig:PZAxsd122}, upper panel, we report the pressure as a function of the baryon density for different values of $y$. The continuous lines correspond to the Gibbs construction in the region of instability of the EOS. For this choice of parameters, the binodal section (lower panel) differs substantially with respect to the previous cases and two separate regions of instability are present. The first one extends at lower pressure and it is present only for small value of $y$, where both mechanical and diffusive instabilities are present. Let us observe that in this lower region, for $y>y_{CP}$, the system goes to a retrograde phase transition likewise to the liquid-gas phase transition.
The upper region of instability extends at greater pressures and higher values of $y$, where mechanical and diffusive instabilities are both present. Also in this second region, on the left of the CP, a retrograde phase transition can occur. However, in this particular case the system is already in a $\Delta$-dominant phase and, at the end of the mixed phase, in which $\Delta$-isobars are partially converted into nucleons, it quickly returns to the $\Delta$-matter phase.

\begin{figure}
\begin{center}
\resizebox{0.45\textwidth}{!}{%
\includegraphics{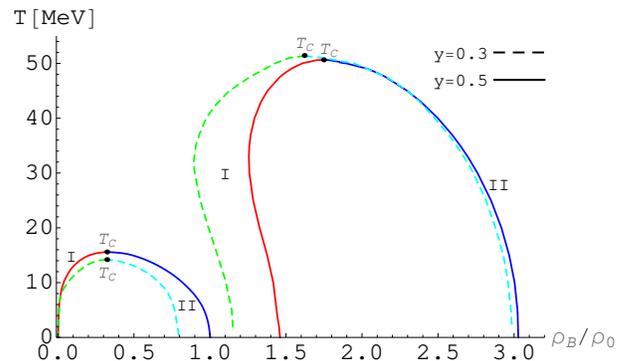}
} \caption{(Color online) Phase diagram of the liquid-gas and the nucleon-$\Delta$ matter phase transition for $y=0.3$ (dashed curves) and $y=0.5$ (continuous curves). The lines labeled with I and II, delimitate the first and second critical densities of the coexistence regions, respectively.} \label{fig:T_rhobtot}
\end{center}
\end{figure}

Finally, in Fig. \ref{fig:T_rhobtot}, we report the phase diagram with in evidence the coexistence regions of the liquid-gas and the nucleon-$\Delta$ matter phase transition for $y=0.3$ and $0.5$ ($x_{\sigma \Delta}=1.3$). The two coexistence regions are well separated and the features of the two phase transitions differ significantly. In fact, for the liquid-gas transition, asymmetric nuclear matter implies a reduction of the second critical density and of the critical temperature $T_c$. Contrariwise, for the $\Delta$-dominant phase transition, we have a slight increase of the critical temperature and a significant reduction of the first critical density. In particular at moderate temperatures ($T\approx 30 \div 40$ MeV), the system begins the mixed phase at a baryon density of the order of $\rho_0$. This behavior could be phenomenologically relevant in order to identify such a phase transition in heavy ion collision experiments.

\section{Conclusions}\label{conclusion}

The main goal of this work is to show the possible existence of chemical and mechanical instability at finite temperature and dense nuclear matter. We have studied the relativistic nuclear EOS with the inclusion of $\Delta$-isobars and by requiring global conservation of baryon and electric charge numbers.
Similarly to the liquid-gas phase transition in a warm and low density nuclear matter, a nucleon-$\Delta$ matter phase transition also can occur at higher temperatures and densities ($T\le 50$ MeV, $\rho\approx 1\div 3\,\rho_0$). We have shown that
for asymmetric nuclear matter both mechanical and chemical instabilities take place. The latter plays a crucial role in the characterization of the phase transition and can also imply very low values of the electric charge fraction $y$ during the mixed phase region.

The nucleon-$\Delta$ matter phase transition depends significantly on the value of the $\sigma$-$\Delta$ coupling constant and we  have seen that only for a limited range of the possible (physical) $x_{\sigma\Delta}$ couplings the presence of thermodynamic instabilities may become relevant from a phenomenological point of view.

Whether metastable $\Delta$-excited nuclear matter exists is still a controversial issue because little is actually known about the $\Delta$-coupling constants with the scalar and vector mesons, even if QCD finite-density sum rule results predict a larger net attraction for a $\Delta$-isobar than for a nucleon in the nuclear medium \cite{jin}.
Although we have seen that instabilities are already present for $x_{\sigma \Delta}>1$, they become phenomenologically more relevant at greater values of $x_{\sigma \Delta}$, involving larger region of mixed phase and greater values of the electric charge fraction.

The analysis of the instability regions with different $\Delta$-coupling constants turns out to be not trivial from the numerical point of view, especially at lower values of $x_{\sigma \Delta}$ where a complex structure of the mixed phase can be formed. For example, we have shown that, in the case of $x_{\sigma \Delta}=1.22$,
two separate regions of instability are present. Moreover, the case $x_{\sigma \Delta}=1.3$ has been studied for different values of temperature and we have seen that in asymmetric nuclear matter the mixed phase transition involves a large range of baryon densities.

Similarly to the liquid-gas phase transition, the nucleonic and the $\Delta$-matter phase have a different electric charge fraction in the mixed phase. The electric charge fraction in the nucleonic  phase reflects a system with higher values of $y$ than the $\Delta$-matter phase.
In the liquid-gas phase transition, the process of producing a larger neutron excess in the gas phase is referred to as isospin fractionation \cite{xu,das,baran}. A similar effects can occur in the nucleon-$\Delta$ matter phase transition due essentially to a $\Delta^{-}$ excess in the $\Delta$-matter phase with lower values of $y$. As already observed, due to the uncertainty on the meson-$\Delta$ coupling constants, we have not considered in this investigation the coupling of the $\Delta$ with the isovector $\rho$-meson field, because this is much less explored in the literature. We have verified that the presence of a $\rho-\Delta$ coupling could further increase the isospin asymmetry in the mixed phase and lower the critical temperature of the nucleon-$\Delta$ matter phase transition.

In this context it is proper to observe that Coulomb interaction and finite size effects, not considered in this study, can significantly alter the structure of the phase transition. Moreover, as already observed, it is proper to remember that our effective EOS cannot incorporate the complex $\pi N\Delta$ dynamics and it would be very interesting to investigate the presence of thermodynamic instabilities in the framework of a more realistic chiral hadronic EOS. Taking also into account the large uncertainty on the possible values of $\Delta$-meson field couplings, it would be prudent to note the pedagogical character of this preliminary study.

Many effects discussed in this paper may be more evident at low values of $y$, obtainable, in principle, with radioactive ion beam facilities. On the other hand, it is rather unlikely, at least in the near future, that neutron rich nuclei can be accelerated to energies larger than a few GeV per nucleon. However, some precursor signals of the considered instabilities could be observed even in collisions of stable nuclei at intermediate energies. For example, in Ref. \cite{ditoro2006}, the simulation of the reaction $^{238}$U+$^{238}$U (average $y=0.39$), at 1 $A$ GeV and semicentral impact parameter $b=7$ fm, shows that rather exotic nuclear matter can be formed in a transient time of the order of 10 fm/$c$, with a baryon density up to $3\,\rho_0$, $T\le 50\div 60$ MeV and $y\approx 0.35\div 0.40$. Such conditions would agree fully with the results for the nucleon-$\Delta$ mixed phase region (see Fig. \ref{fig:T_rhobtot}).

A possible signature of the nucleon-$\Delta$ matter phase transition could be found via observables particularly sensitive to the expected different isospin content of the two phases. For example, at the AGS energies, the $\Delta$-resonance was predicted to be the dominant source for pions of small transverse momenta \cite{hofmann}. In this case, an increase of the negative pions $\pi^{-}$ of small trasverse momenta at a greater asymmetry of the beam could be a good indicator of a $\Delta$ isospin fractionation effect.


\end{document}